\documentclass[
 reprint,         % two-column
 superscriptaddress,
 amsmath,amssymb,
 aps,
 prl,             % journal
 floatfix
]{revtex4-2}

\usepackage{graphicx}
\usepackage{dcolumn}
\usepackage{bm}
\usepackage[colorlinks=true,allcolors=blue]{hyperref}
\usepackage{siunitx} % nice units: \SI{3.2}{\micro\meter}

\begin{document}
\title{Engineering Rogue Waves via Multimode Interactions in Integrated Waveguides}

\author{Gülsüm Yaren Durdu}
\affiliation{Department of Electrical and Electronics Engineering, Koç University, Istanbul, Türkiye}

\author{Azka Maula Iskandar Muda}
\affiliation{Department of Electrical and Electronics Engineering, Koç University, Istanbul, Türkiye}

\author{Uğur Teğin}
\email{utegin@ku.edu.tr}
\affiliation{Department of Electrical and Electronics Engineering, Koç University, Istanbul, Türkiye}

\begin{abstract}
We explore rogue wave formation in multimode silicon nitride (Si$_3$N$_4$) waveguides with multimode nonlinear Schrödinger equation-based simulations. Pure fundamental-mode excitation produces smooth propagation without extreme events, whereas higher-order modes or multimode superpositions yield heavy-tailed statistics with bursts exceeding the $8\sigma$ threshold. These results reveal that rogue wave generation in integrated waveguides is controlled not only by material properties such as nonlinearity and dispersion but also by modal excitation and intermodal nonlinear interactions. Our results identify modal control as a new degree of freedom for engineering extreme spatiotemporal events on photonic chips, with implications for on-chip supercontinuum generation, frequency combs, and nonlinear wave management.
\end{abstract}

\maketitle
% --- Main text (no numbered sections in PRL; use paragraphs) ---
\paragraph{Introduction.} 
For centuries, sailors recounted encounters with giant waves that appeared suddenly on the high seas, destroyed vessels, and vanished without a trace. These so-called rogue waves remained part of maritime lore until January 1, 1995, when an instrument on the Draupner platform in the North Sea recorded a single wave of unprecedented height, transforming a centuries-old legend into a subject of scientific investigation \cite{1,2}. In oceanography, rogue waves are defined as surface waves whose height exceeds eight times the standard deviation of the sea surface elevation, or equivalently, twice the significant wave height, thus linking their existence to extreme statistical outliers \cite{3}. Since then, the concept of rogue waves has expanded well beyond the ocean. Analogous extreme events have been identified in acoustic, thermal, and financial systems \cite{4}.

The discovery of optical rogue waves, rare, high-intensity fluctuations observed during supercontinuum generation in optical fibers, became possible in 2007 with advances in real-time spectral measurements \cite{5}. These temporally localized events, driven by noise-seeded modulation instability and soliton collisions, revealed the sensitivity of nonlinear fiber dynamics to initial conditions and stimulated extensive studies on the interplay between noise, nonlinearity, and dispersion \cite{6,7,8}. Spatial analogues were later reported in filamentation experiments and highly multimodal fibers acting as complex scattering media \cite{9,10}.

The diversity of fiber-based platforms has revealed a rich landscape of mechanisms, and rogue waves were also demonstrated in dissipative systems such as mode-locked fiber lasers, where the interplay of gain, loss, and nonlinear pulse shaping produces chaotic multi-pulse regimes with intensity spikes far exceeding Gaussian statistics \cite{11,12,13,14}. Recently, real-time observation of rogue waves in spatiotemporally mode-locked fiber lasers was reported using compressed ultrafast photography, showing that multimode cavities can sustain rare, extreme-intensity pulses shaped by a nonlinear attractor in graded-index fibers \cite{15}. 

Recent advances in integrated nonlinear photonics, particularly with low-loss Si$_3$N$_4$ waveguides, have established this platform as one of the most mature and versatile in the field \cite{16,17,18}. Over the past decade, Si$_3$N$_4$ technology has achieved record-low propagation losses below 0.1 dB/m and demonstrated wafer-scale fabrication with high yield, enabling reproducible and large-volume production of photonic circuits. Combined with its high Kerr nonlinearity, broad transparency from the visible to the mid-infrared, and precise dispersion engineering, Si$_3$N$_4$ has powered breakthroughs in microresonator frequency combs, octave-spanning supercontinuum generation, ultrafast pulse sources, and emerging applications such as quantum photonics \cite{19,20}.

While early efforts focused on single-mode devices, recent studies have extended Si$_3$N$_4$ to multimode regimes, revealing a new landscape of spatiotemporal nonlinear dynamics. Intermodal four-wave mixing, multimode soliton formation, dispersive wave generation, and supercontinuum broadening have all been demonstrated, with input mode engineering emerging as a powerful tool for tailoring nonlinear interactions \cite{21,22,23}. These developments position multimode Si$_3$N$_4$ waveguides as a uniquely capable platform for studying and controlling extreme spatiotemporal events at the chip scale, bridging the gap between fiber-based rogue wave physics and integrated photonics.

In this Letter, we investigate rogue wave formation in multimode Si$_3$N$_4$ waveguides and demonstrate that it can be enhanced through initial modal excitation. We study pulse propagation across fundamental and higher-order transverse electric modes using a multimode nonlinear Schrödinger equation (MM-NLSE) framework with stochastic noise to seed modulation instability. Our large-scale simulations reveal that pure fundamental-mode excitation leads to smooth, deterministic evolution. In contrast, higher-order modes or multimode superpositions induce strong intermodal coupling, spectral broadening, and heavy-tailed intensity statistics with events exceeding the $8\sigma$ threshold. These findings establish modal excitation as a powerful degree of freedom for controlling extreme spatiotemporal events in integrated nonlinear waveguides. It opens new opportunities for on-chip studies of rare-event dynamics and their potential applications.

\paragraph{Method.}
We modeled pulse propagation in multimode Si$_3$N$_4$ waveguides by solving the multi-mode nonlinear Schrödinger equation (MM-NLSE) accounting for dispersion, self- and cross-phase modulation, and intermodal four-wave mixing \cite{23,24}.
\begin{equation} \label{eq:1}
\begin{split}
\frac{\partial A_p}{\partial z} = & \ i\left(\beta_0^{(p)}-\beta_0\right)A_p(z,t) - \left(\beta_1^{(p)}-\beta_1\right)\frac{\partial A_p(z,t)}{\partial t} \\
& + i\sum_{n \ge 2}\frac{\beta_n^{(p)}}{n!}\left(i\frac{\partial}{\partial t}\right)^n A_p(z,t) \\
& + i\frac{n_2\omega_0}{c}\sum_{l,m,n} \eta_{plmn}(\omega_0)A_l(z,t)A_m(z,t)A_n^*(z,t)
\end{split}
\end{equation}
where $\eta_{plmn}$ is nonlinear coupling coefficient, $n_2$ is Kerr coefficient, $\omega_0$ is the central frequency, $\beta_n^{(p)}$ is the dispersion value of the corresponding mode. The nonlinear coupling tensor, which governs the intermodal coupling coefficients between the modes, is determined from the spatial overlap integral. The fourth-order Runge-Kutta in the interaction picture algorithm is utilized to study light propagation with high accuracy.  Pulses with a peak power of 100 kW, a wavelength of 1030 nm, and a duration of 250 fs are launched into different mode configurations in the waveguide. Stochastic noise was added to the input beam to seed modulation instability to generate rare, extreme-intensity rogue waves. Noisy input is
\begin{equation} \label{eq:3}
A_p(0,t) = c_p A_0(t)\left[1 + \delta n(t)\right]
\end{equation}
where $\delta$ is the noise fraction of the noise strength, $n(t)$ is an independent standard normal random variable at each sample time, $c_p$ is the normalized modal coefficient for mode p, and $A_0(t)$ is the noiseless temporal envelope.

\begin{figure}[t!]
    \centering
    \includegraphics[width=\columnwidth]{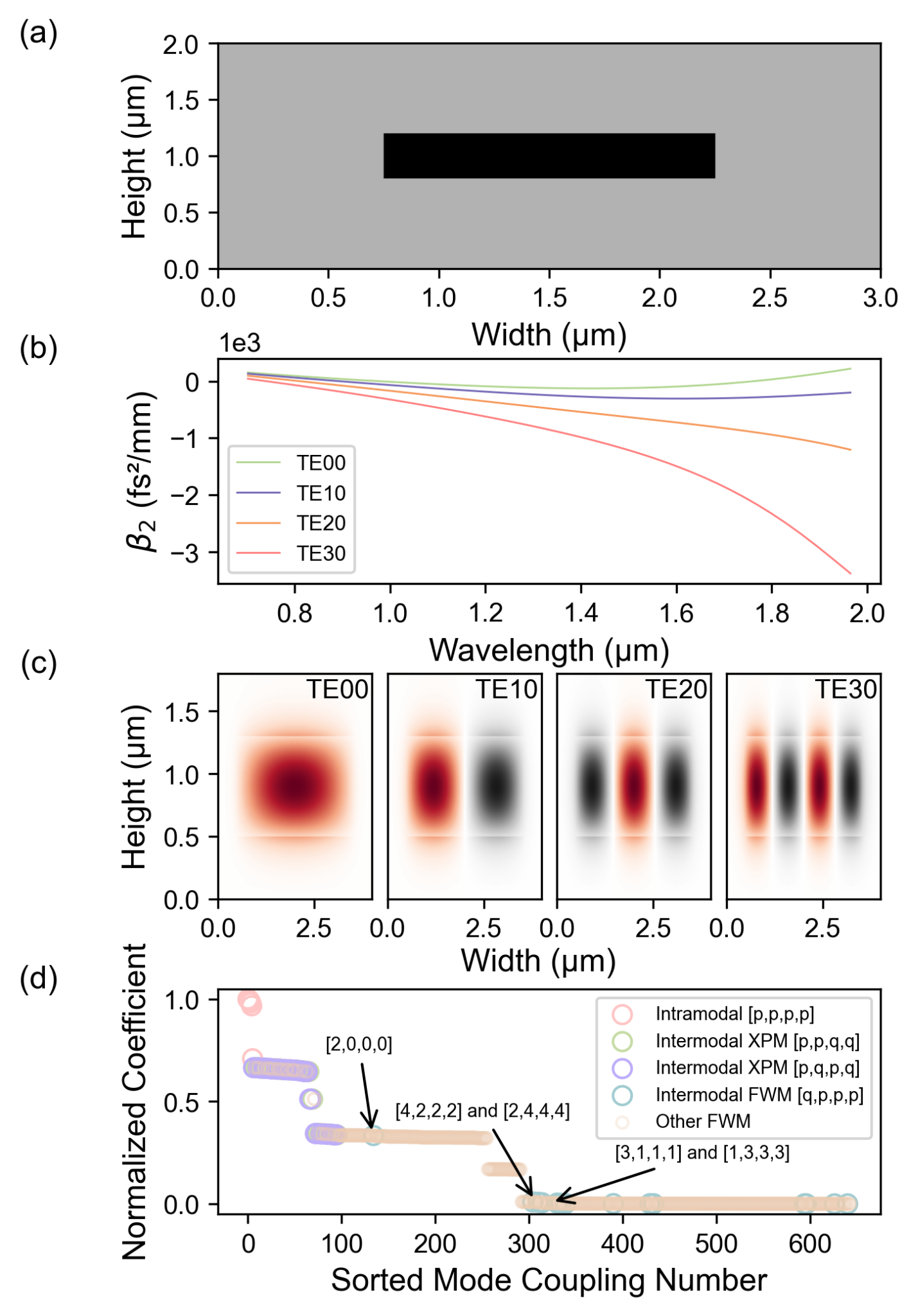}
    \caption{Properties of waveguide (a) Waveguide cross-section (b) Group velocity dispersion (GVD) of four guided modes. (c) Mode profiles of TE00, TE10,TE20,TE30 modes (d) Normalized nonlinear coupling coefficients for intermodal and intramodal interactions}
    \label{fig:1}
\end{figure}

\begin{figure}[t!]
    \centering
    \includegraphics[width=\columnwidth]{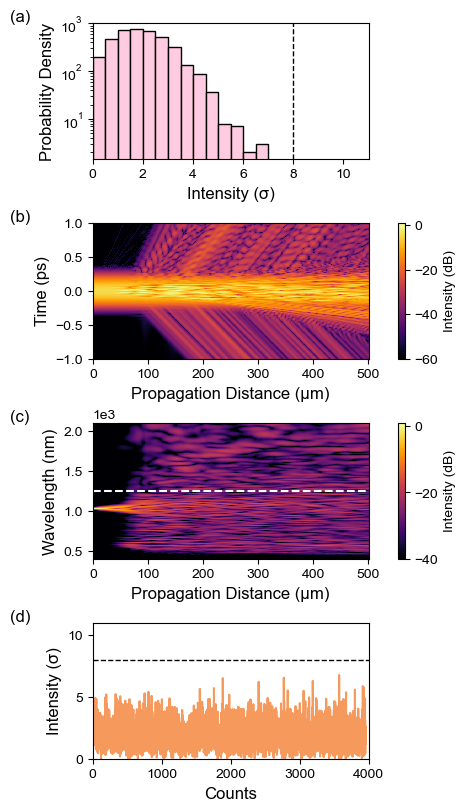}
    \caption{Fundamental mode excitation (a) Probability density function of normalized intensity. (b)Temporal evolution of the field. (c) Spectral evolution of the field (d) Peak intensity traces over 4000 simulations.}
    \label{fig:2}
\end{figure}

The statistics were evaluated within the red-shifted spectral window between 1250 and 2100 nm to identify extreme events. Using a numerical eigenvalue mode solver, the four modes' propagation constants and higher-order dispersion coefficients are extracted, capturing group-velocity dispersion, group velocity mismatch, and additional higher-order dispersion terms. Nonlinear effects-including self-phase modulation (SPM), cross-phase modulation (XPM), and intermodal four-wave mixing (FWM) - were obtained from mode overlap integrals and intrinsic material properties of Si$_3$N$_4$ given in Fig. \ref{fig:1}. 

The cross-section of 500 $\mu$m x 3 $\mu$m x 800 nm waveguide in Fig 1(a). Supports four transverse electric modes, whose dispersion parameters are shown in Fig 1(b). The fundamental mode (TE00) demonstrates a slowly varying dispersion profile, while higher modes show increasingly decreasing $\beta_2$ resulting in an anomalous dispersion regime favorable for phase-matched interactions and growth of modulation instability. The transverse mode profiles in Fig 1(c). display spatial overlap between the modes, which is quantified in Fig. 1(d). While intramodal interactions are sufficient, several intermodal interactions, such as TE00 and TE20, are pronounceable, contributing to the overall nonlinear dynamics of the waveguide.

\paragraph{Results.}
Our numerical simulations reveal that rogue wave generation in multimode waveguides critically depends on the initial modal excitation. To seed the modulation instability (MI) that precipitates extreme events, we introduced a 1\% stochastic noise level to the Gaussian input pulse. We performed 4000 simulations for each condition to ensure robust statistical analysis.

We observe that excitation of the fundamental TE$_{00}$ mode alone is insufficient to trigger rogue events. The peak intensity's probability distribution function (PDF) follows a clear exponential decay, with no events exceeding the rogue threshold of 8$\sigma$ as illustrated in Fig. 2(a). Temporal and spectral propagations of the pulse with the highest intensity above 1250 nm are presented in Figs. 2(b) and 2(c). Although they present a highly dispersive nonlinear propagation, a significant intensity build-up is not observed. Consequently, the peak intensity trace across all 4000 simulations remains well below the rogue threshold (see Fig. 2(d)). This suggests that intramodal nonlinear effects alone do not drive the system into the extreme event regime under these initial conditions.

\begin{figure}[t]
    \centering
    \includegraphics[width=\columnwidth]{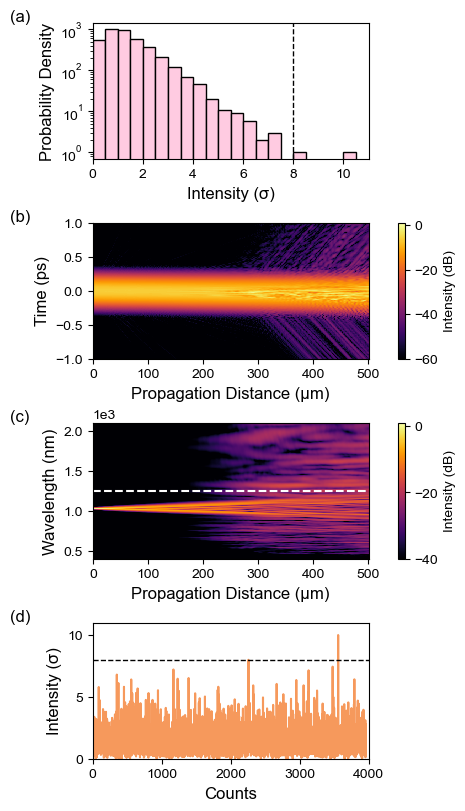}
    \caption{Equal excitation of TE00 and TE20 modes (a) Probability density function of normalized intensity. (b)Temporal evolution of the field. (c) Spectral evolution of the field (d) Peak intensity traces over 4000 simulations.}
    \label{fig:3}
\end{figure}

In stark contrast, launching an equal superposition of the TE${00}$ and TE${20}$ modes facilitates robust rogue wave formation. The PDF develops a pronounced, heavy-tailed distribution, with multiple events exceeding 8$\sigma$ [Fig. 3(a)]. Temporal and spectral propagations of the pulse with more than 10 $\sigma$ intensity above 1250 nm are presented in Figs. 3(b) and 3(c). The symmetric nature of dispersive action in the time domain indicates that the dominant nature of nonlinear propagation is four-wave mixing-based. These features are clear signatures of a highly nonlinear process driven by strong intermodal interactions, which trigger the MI necessary for rogue wave generation.

\begin{figure}[t]
    \centering
    \includegraphics[width=\columnwidth]{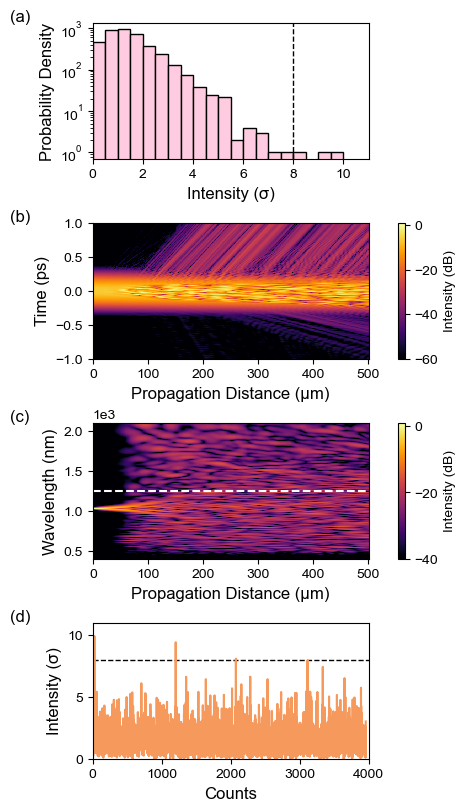}
    \caption{Only TE10 mode excitation (a) Probability density function of normalized intensity. (b)Temporal evolution of the field. (c) Spectral evolution of the field (d) Peak intensity traces over 4000 simulations.}
    \label{fig:4}
\end{figure}

In our studies, we observe that selectively exciting a single higher-order mode, such as the TE$_{10}$ mode, also generates rogue waves effectively. This configuration yields a heavy-tailed PDF with a significantly high number of events beyond the 8$\sigma$ threshold as demonstrated in Fig. 4(a). Temporal and spectral propagations of the pulse around 10 $\sigma$ intensity above 1250 nm are presented in Figs. 4(b) and 4(c). The statistical trace of peak intensities confirms the presence of numerous rogue events across the simulation ensemble (see Fig. 4(d)).

\paragraph{Discussion and Conclusion.}
Our results reveal a clear link between initial modal excitation and the onset of rogue waves in multimode Si$_3$N$4$ waveguides. Two distinct regimes emerge: a stable propagation regime dominated by the fundamental TE${00}$ mode, and a regime prone to rogue waves triggered by higher-order or multimode excitation. While extreme events can, in principle, arise in single-mode systems, our simulations show that under moderate input conditions the fundamental mode alone remains in a stable regime, with rare-event statistics suppressed below the $8\sigma$ threshold.

The dramatic enhancement of the probability of rogue waves with multi-mode excitation originates from the rich intermodal non-linear dynamics. Spatiotemporal beating between modes introduces periodic intensity modulation that seeds modulation instability (MI) far more efficiently than stochastic noise alone. Moreover, intramodal and intermodal dispersion coexistence provides additional phase-matching pathways for energy localization, accelerating the growth of high-amplitude, localized peaks. By contrast, the absence of intermodal coupling in the single-mode case limits the system to weak MI gain, preventing the formation of extreme events under the same conditions.

In summary, we have demonstrated that the initial modal composition of the input pulse serves as a powerful control parameter for nonlinear dynamics in integrated waveguides. Exciting either a single higher-order TE${10}$ mode or a superposition of TE${00}$ and TE$_{20}$ modes drives the system into a regime of strong MI gain and heavy-tailed statistics. In contrast, pure fundamental-mode excitation maintains stable, near-Gaussian dynamics.

These findings establish intermodal interactions as a practical tool for tailoring nonlinear instabilities in integrated photonic platforms. Beyond fundamental studies of optical rogue waves, such control could be exploited for on-chip supercontinuum sources \cite{23}, frequency comb generation, and other nonlinear photonic applications enabled by CMOS-compatible Si$_3$N$_4$ technology.

% --- Acknowledgments ---
\begin{acknowledgments}
Funding: Optica Foundation Challenge Prize.
\end{acknowledgments}

% --- Data/Code availability (optional but encouraged) ---
%\paragraph{Data availability.}
%Data and code are available at \url{https://...}.

% --- Bibliography ---
\bibliographystyle{apsrev4-2}
\bibliography{refs}

% --- Supplementary Material (optional, separate file for submission) ---
% \onecolumngrid
% \appendix
% \section*{Supplemental Material}
% Details that support the main claims without being essential for reading flow.

\end{document}